\documentclass[11pt]{article}

\begin{document}
\vspace{25pt}
\title{\bf{Berry Phase for Systems with Angular Momenta in Electric and Magnetic Fields}}
\vspace{5pt}
\author{\bf{K.J.B. Ghosh, D. De Munshi and B. Dutta-Roy }} \maketitle
\begin{center}
\title{\bf{Abstract}}
\end{center}

The Berry phase for a variety of systems comprising of two angular
momenta is discussed. These include the electron and proton in the
ground state of the hydrogen atom (taking into account the
hyperfine interaction), the positronium atom ($^{3}S_{1}$ and
$^{1}S_{0}$ states), the $\mu^{\pm}e^{\mp}$ and $\mu^{+}\mu^{-}$
bound states, the spin and orbital angular momenta of a single
electron, $\mu^{-}$ capture by the deuteron in an external magnetic
field, etc. Though the time scales involved, the underlying
intrinsic Hamiltonians are quite different, as also the possible
experimental probes, the geometric nature of the results for the
Berry phase due to a time varying externally imposed magnetic field
is found to be quite robust. Some indications are also put forward
as to the possible interesting studies with time varying electric
fields. The objective of this work is an attempt to broaden the
scope of such studies in both the experimental and theoretical
directions.
\newpage
\section{Introduction}\label {section1}
The effect of Berry phase can be found in a variety of systems such
as in optics, especially polarization of light (ref.
\cite{panchpaper},\cite{chiopaperI},\cite{chiopaperII}), molecular
spectroscopy \cite{herzpaper}, atom-molecule scattering
\cite{meadpaper},nuclear quadrupole resonance \cite{tyckopaper},
quantum Hall effect \cite{arovaspaper} and many other fields ranging
from optics to condensed matter physics to atomic and molecular
physics. Hence understanding of Berry phase enables one to
comprehend a variety of phenomenon observed in diverse fields of
study.\\
Even though the concept of Berry phase \cite{berrypaper} arose most naturally with the physical setting of a state of a quantum system with a time dependent Hamiltonian, where the adiabatic approximation can be applied, whereby the state clings to the instantaneous eigenstate of the changing Hamiltonian, nevertheless it emerges as an important intrinsic geometric property of the state in question, in the space of parameters that characterize the time dependence of the Hamiltonian. The Berry phase $\gamma_n(t)$ manifests itself as an extra phase factor in the eigenfunction of the instantaneous Hamiltonian (with eigenvalue $E_{n}(t)$ ) over and above the familiar dynamical phase $-\frac{1}{\hbar}\int_{0}^{t}E_{n}(t')dt'$. Thus even if the change in the Hamiltonian is cyclic so that $H(t=T)=H(t=0)$ a non-integrable phase factor $\gamma_{n}(T)$ for the eigenfunction remains as an anholonomy. Moreover $\gamma_{n}(T)$ can be expressed in terms of the solid angle (or its generalization) of the path (circuit) executed in parameter space (through time $t=0$ to $T$ ) subtended at the point in the parameter space where the degeneracies lifted by the time dependent part of the Hamiltonian are restored (known as the "diabolical point"). This is the geometric aspect of the Berry phase. In analogy with the phase that the wavefunction of a charged particle acquires upon execution of a closed path in a magnetic field, which is related to the flux enclosed by that circuit, one can also associate an underlying gauge field with the Berry phase( ref \cite{wilzchekpaper},\cite{ivanovpaper}). Depending on whether the state (labeled here by $n$) of the system under consideration being 'transported', so to say, is non-degenerate or degenerate, the under lying gauge field is Abelian or Non-Abelian.\\
For concreteness (and also for relevance to the present study), we
focus on a particle (or system of particles) with angular momenta
and associated magnetic moments (with corresponding gyromagnetic
ratios) in an external time dependent magnetic field
$\overrightarrow{B}(t)$. Thus the Hamiltonian governing the time
dependence of the system is itself time-dependent, as encoded in the
three dimensional parameter space (in this case) describing the
magnetic field. Furthermore, let us for instance allow the magnetic
field to vary with time in such a manner as to keep its magnitude
and polar angle fixed but its azimuth ($\phi$) increasing uniformly
with time with some angular frequency ($\omega$) such that
$\phi=\omega t$ and
\begin{equation}
\overrightarrow{B}=B_{0}(sin\theta cos\phi \hat{i} + sin \theta sin
\phi \hat{j} + cos \theta \hat{k}) \label{magfield}
\end{equation}
Thus the magnetic field vector, in the parameter space traces out in time a cone with a semi-vertical angle $\theta$. The Berry phase was found in the case of a spin half particle, to be determined by the solid angle $\Omega_{c}$ subtended at $\overrightarrow{B}=0$ (the point in the parameter space where the Zeeman splitting vanishes) by the closed curve traced out in that space by the tip of $\overrightarrow{B}$ after one cycle of evolution viz. $T=\frac{2 \pi}{\omega}$. For the particular case at hand (Eq. \ref{magfield}), this solid angle is given by $\Omega_{c} = 2 \pi (1-cos\theta)$ and for the states of the particle of spin half, the Berry phase is given by $\gamma_{m} (C)= -m \Omega_{c}$ with $m=\pm \frac{1}{2}$ for the two states. Moreover, the generalization of the results to the case of the arbitrary angular momentum $j$ is readily accomplished and simply yields $\gamma_{m} (C)= -m \Omega_{C}$ with $m=-j.....+j$.\\
More recently there have been papers (ref
\cite{gqzhupaper},\cite{dmtongpaper},\cite{xchangtangpaper}) devoted
to the Berry phase under the conditions similar to what has been
described above, for a system of two identical spin half particles
keeping in mind relevance to the entangled states in context the of
Quantum Information Theory. One of the points we wish to emphasize
in the present study is the robustness of Berry's result not only
for the case of two identical spin one-half particles but for a more
general scenario where the particles are non-identical and carry
different gyromagnetic ratios. Indeed the results seem to be more
group theoretical than geometrical.The main difference in the
generic structure of the two situations arises from the fact that in
the case of identical particles, the interaction Hamiltonian with
the magnetic field is symmetric under the exchange of the two spins
and hence the magnetic field is incapable of causing ,through its
interaction, any admixture between spin singlet (antisymmetric under
spin exchange) and spin triplet (symmetric under spin exchange)
states. Thus the basic nature of the interaction Hamiltonian in the
case of non-identical particles is quite different. The second
aspect of our study pertains to the variety of such systems of two
spin one-half particles, representing a vast range of energy and
timescales, as also the diversity of the underlying experimental
techniques and probes for investigation. Thus, for example, one may
consider the positronium atom composed of the electron and the
positron in the two lowest $^{1}S_{0}$ (para) and $^{3}S_{1}$
(ortho) states, each having their own distinctive decay modes into
two and three gammas with lifetimes $~10^{-10}$ and $10^{-7}$
seconds respectively. This system, in the present context, is the
very antithesis of the case of two identical electrons, because here
the interaction with the magnetic field is antisymmetric under the
exchange of spins because the gyromagnetic ratios of the electron
and positron are equal in magnitude and opposite in sign. Another
example at the other extreme of the time and energy scale are the
lowest hyperfine interaction split triplet and singlet states of the
hydrogen atom where the transition between the former and the latter
states gives rise to the famous 21 cm line of the electromagnetic
spectrum and has a lifetime of approximately ten million years!
Another amusing set of examples involve the muon (which has a
lifetime of $2.2$ microseconds), such as the $\mu^{+}e^{-}$,$\mu^{-}
e^{+}$ and $\mu^{+}\mu{-}$ bound states. Here an interesting
experimentally exploitable feature is the fact that through its
parity violating weak decay, the muon self-analyzes its state of
polarization, through the angular distribution of its decay
electrons.

\section{System with Two Angular Momenta in a Time Varying Magnetic Field}
\subsection{Systems with Two Spin One-Half Particles}
The generic Hamiltonian under study is
\begin{equation}
H=G(\overrightarrow{S_{1}}\cdot\overrightarrow{S_{2}})-\mu_{B}\overrightarrow{B}(
t).(g_{1}\overrightarrow{S_{1}}+g_{2}\overrightarrow{S_{2}})
\label{hamil2}
\end{equation}
where $G$ is the strength of the 'hyperfine' interaction between the
two spins, $\mu_{B}=\frac{e\hbar}{2mc}$ is the Bohr magneton with
$m=m_{e}$ for electron, $m=m_{p}$ for proton and $m=m_{\mu}$ for
muon and $g_{1}$ and $g_{2}$ are the gyromagnetic ratios of the two
particles involved. X.C. Tang $et.al.$ (ref \cite{xchangtangpaper})
confining their attention to the system of two identical particles
(the special case with $g_{1}=g_{2}$ ), in the rotating magnetic
field (Eq.\ref{magfield}) found that here too Berry's result for the
phase was obtained if the quantum number (there for a single
particle) is replaced by that of the respective entangled state. In
the more general scenario that we are considering, it is convenient
to introduce the total angular momentum
$\overrightarrow{S}=\overrightarrow{S_{1}}+\overrightarrow{S_{2}}$
and the difference angular momentum $\overrightarrow{\Delta
S}=\overrightarrow{S_{1}}-\overrightarrow{S_{2}}$ and to rewrite the
Hamiltonian as
\begin{eqnarray}
\nonumber H=\frac{G}{2}(S^2-{S_{1}}^2-{S_{2}}^2)-\mu_{B} B_{0} g_{+}
[\frac{1}{2} sin{\theta} (S_{+}e^{-i \phi}+S_{-} e^{-i \phi}) +
cos{\theta}
S_{z}]\\
 -\mu_{B} B_{0} g_{-}[\Delta S_{x} sin{\theta} cos{\phi}+\Delta S_{y}
sin{\theta} sin{\phi}+ \Delta S_{z} cos{\theta}]
 \label{finhamil}
\end{eqnarray}
where we have introduced $g_{\pm}=\frac{1}{2} (g_{1} \pm g_{2})$ and the ladder operators $S_{\pm}=S_{x} \pm iS_{y}$. Since we have two spin one-half particles a natural choice of the basis states are the singlet ($|0,0>$) and the triplet ($|1,1>,|1,0>,|1,-1>$) states. If $g_{-}=0$ ( the case of identical particles) the matrix elements of the Hamiltonian connecting the singlet and the triplet states are zero. In general $g_{-}\neq0$ (the case we are considering) and the structure of the Hamiltonian is much more interesting, since coupling between the singlet and the triplet states are permitted.\\
The following equations give the eigenvalues and eigenvectors of $H$
respectively
\begin{equation}
E_{1}=\frac{\eta}{2}+\gamma_{+} \\
\nonumber
\end{equation}
\begin{equation}
  |n_{1}>=\frac{1}{\sqrt{2}}sin\theta e^{i\phi} |1,0> + \frac{1}{2}
(1-cos\theta) e^{2i\phi} |1,-1> + \frac{1}{2}(1+cos\theta) |1,1>  \\
  \end{equation}
\\
  \begin{equation}
  \nonumber E_{2}=\frac{\eta}{2}-\gamma_{+} \\
  \end{equation}
  \begin{equation}
  |n_{2}>=\frac{1}{\sqrt{2}}sin\theta e^{-i\phi} |1,0> - \frac{1}{2}
(1-cos\theta) |1,-1> + \frac{1}{2}(1+cos\theta) e^{-2i \phi} |1,1> \\
  \end{equation}
\\
  \begin{equation}
  \nonumber E_{3}= -\frac{\eta}{8}+k \\
  \end{equation}
  \begin{equation}
  |n_{3}>= cos\chi |0,0> -sin\chi cos\theta |1,0> -\frac{1}{\sqrt{2}} sin\chi
sin\theta e^{i\phi} |1,-1> + \frac{1}{\sqrt{2}} sin\chi sin\theta
e^{-i\phi}
|1,1>\\
  \end{equation}
\\
  \begin{equation}
  \nonumber E_{4}= -\frac{\eta}{8}-k \\
  \end{equation}
  \begin{equation}
  |n_{4}>=sin\chi |0,0> +cos\chi cos\theta |1,0> +\frac{1}{\sqrt{2}} cos\chi
sin\theta e^{i\phi} |1,-1> - \frac{1}{\sqrt{2}} cos\chi sin\theta
e^{-i\phi}
|1,1>\\
  \end{equation}
  where,\\
\begin{eqnarray}
  \nonumber \eta=G\hbar \\
  \nonumber \gamma_{\pm}=-\mu_{B} B_{0}\hbar g_{\pm} \\
  \nonumber tan\chi=\frac{\gamma_{-}}{\frac{5\eta}{8}+k} \\
  \nonumber k=\sqrt{(\frac{5\eta}{8})^2+(\gamma_{-})^2}
\end{eqnarray}

The geometric phase for the nth eigenstate is given as-\\
\begin{equation}
\gamma_{n}=i\oint <n(R)|\nabla_{R} n(R)> dR\\
\label{oriberry}
\end{equation}

The circuit integral is in the parameter space where the hamiltonian
is varied adiabatically. In our system the variable parameter is
$\phi$. Hence Eq. \ref{oriberry}
becomes-\\
\begin{equation}
\gamma_{n}=i\int_{0}^{2\pi} <n(\phi)|\frac{\partial}{\partial\phi}
n(\phi)>
d\phi \\
\label{ourberry}
\end{equation}

Applying the above formula to the eigenstates of $H$, we get,
\begin{eqnarray}
  \gamma_{n_{1}}=-2\pi(1-cos\theta) \\
  \gamma_{n_{2}}=2\pi(1-cos\theta) \\
  \gamma_{n_{3}}=0 \\
  \gamma_{n_{4}}=0
\end{eqnarray}

\subsection{Systems with One Spin $1$ and One Spin $\frac{1}{2}$ Angular Momenta}
The system of one spin 1 and one spin half particle is also of
importance in several scenarios. It can describe spin orbit coupling
of electrons in the first excited state of atoms of alkali metals(
or singly charged alkali earth metal ions), as well as hyperfine
interaction between electron in the ground
state of a  deuterium atom and its nucleus.\\

We take the Hamiltonian same as Eq.\ref{finhamil} (in the case where
the spin one corresponds to the orbital angular momenta the
gyromagnetic ratio is $1$). For a system of two angular momenta, one
and one-half, the natural choice of the basis states are
$|\frac{3}{2},\frac{3}{2}>$, $|\frac{3}{2},\frac{1}{2}>$,
$|\frac{3}{2},-\frac{1}{2}>$, and$|\frac{3}{2},-\frac{3}{2}>$, which
we call the quartet states and
$|\frac{1}{2},\frac{1}{2}>$, $|\frac{1}{2},-\frac{1}{2}>$, which we call the doublet states.\\
The following table gives the Berry phase for the eigenstates of the
Hamiltonian (Eq.\ref{finhamil})

\begin{tabular}{|c|c|c|c|}
  \hline
  \\
  Eigenstates & Composition & Energy Eigenvalues & Berry phase \\
  \\
  \hline
  \hline
  \\
  $|n_{1}>$ & Quartet states only & $\frac{\eta}{4}$+
($\frac{3\gamma_{+}}{2}$ + $\frac{\gamma_{-}}{2})$ & $-3\pi (1-cos\theta)$ \\
  \\
  \hline
  \\
  $|n_{2}>$ & Quartet states only & $\frac{\eta}{4}$-
($\frac{3\gamma_{+}}{2}$ + $\frac{\gamma_{-}}{2})$ & $3\pi (1-cos\theta)$ \\
  \\
  \hline
  \\
  $|n_{3}>$ & Quartet and doublet states & $\frac{\eta}{4}$+
($\frac{\gamma_{+}}{2}$ + $\frac{\gamma_{-}}{6})k $& $-\pi (1-cos\theta)$ \\
  \\
  \hline
  \\
  $|n_{4}>$ & Quartet and doublet states & $\frac{\eta}{4}$-
($\frac{\gamma_{+}}{2}$ + $\frac{\gamma_{-}}{6})k $& $-\pi (1-cos\theta)$ \\
  \\
  \hline
  \\
  $|n_{5}>$ & Quartet and doublet states & $-\eta$ + ($\frac{\gamma_{+}}{2}$
+ $\frac{5\gamma_{-}}{6})k $& $\pi (1-cos\theta)$ \\
  \\
  \hline
  \\
  $|n_{6}>$ & Quartet and doublet states & $-\eta$ - ($\frac{\gamma_{+}}{2}$
+ $\frac{5\gamma_{-}}{6})k $& $\pi (1-cos\theta)$ \\
  \\
  \hline
\end{tabular}
\\
where,\\
\begin{eqnarray}
 \nonumber \eta=J\hbar \\
 \nonumber \gamma_{\pm}=-\mu_{B} B_{0}\hbar g_{\pm} \\
 \nonumber \\
 \nonumber k^{2} = \frac{(\gamma_{+} + 3
\gamma_{-})(\gamma_{+}+\gamma_{-})}{(\gamma_{+} + \frac{5
\gamma_{-}}{3})(\gamma_{+} + \frac{\gamma_{-}}{3})} \\
 \nonumber
\end{eqnarray}

\section{ Systems with Electric Quadrupole Moment in a Time Varying Electric Field}
To discuss the electrical analogue of the problem considered so far,
it is suitable to express the electric quadrupole moment operator (
a second rank symmetric tensor) as built up from the angular
momentum
vector, viz.,\\
\begin{equation}
Q_{ik}=\frac{1}{2} (J_{i}J_{k} + J_{k}J_{i})-\frac{1}{3}\delta_{ik}
\overrightarrow{J}.\overrightarrow{J} \label{quadoper}
\end{equation}
As such the $Q_{z^{'}z^{'}}$ component (with $z^{'}$ the time
dependent direction along which the $z^{'}$ component of the
electric field has a gradient $q_{0}$) of this tensor is given by
\begin{equation}
Q_{z^{'}{z^{'}}}=J_{z^{'}}^2-\frac{1}{3} \overrightarrow{J}^2
\label{quadz}
\end{equation}
which with\\
\begin{equation}
J_{z^{'}}=J_{x} sin\theta cos\phi + J_{y} sin\theta sin\phi + J_{z}
cos\theta \label{timedepz}
\end{equation}
will, with $\theta$ fixed and $\phi=\omega t$, provide us with the electric correspondence of the magnetic field vector executing
gyration along a cone. Thus we have

\begin{eqnarray}
Q_{z^{'}z^{'}}=J_{x}^2 (sin^2\theta cos^2\phi-\frac{1}{3})+ J_{y}^2
(sin^2\theta sin^2\phi-\frac{1}{3})\\
\nonumber + J_{z}^2 (cos^2\theta-\frac{1}{3})
+(J_{x}J_{y}+J_{y}J_{x})sin^2\theta cos\phi
sin\phi\\
\nonumber
+(J_{y}J_{z}+J_{z}J_{y}) sin\theta cos\theta
sin\phi+(J_{z}J_{x}+J_{x}J_{x}) sin\theta cos\theta cos\phi
\label{fullquad}
\end{eqnarray}

The Hamiltonian of the interaction of the quadrupole moment of the
system with the 'rotating' electric field gradient is (taking into
account factors arising from the Taylor expansion of the electric
potential and the conventional definition of the electric quadrupole
moment $Q$) given by
\begin{equation}
H_{int}=(\frac{e^2 q_{0}
Q}{12})(J^2_{z^{'}}-\frac{1}{3}\overrightarrow{J}^2)
\label{quadhamil}
\end{equation}
Let us for concreteness consider the simplest example of a particle
with spin $j=1$ (for instance the deuteron). The eigenstates of
$J_{z^{'}}$ belonging to eigenvalues $+1$, $-1$ and $0$, written in
the basis in which $J_{z}$ is diagonal are
\begin{equation}
|\chi_{+1}>=\frac{1}{2}(1+cos\theta) e^{(-i \phi)} |1,+1> +
\frac{1}{\sqrt{2}}sin\theta |1,0>+\frac{1}{2}(1-cos\theta)e^{+i\phi}|1,-1>\\
\nonumber
\end{equation}

\begin{equation}
|\chi_{-1}>=\frac{1}{2}(1-cos\theta) e^{(-i \phi)} |1,+1> -
\frac{1}{\sqrt{2}}sin\theta |1,0>+\frac{1}{2}(1+cos\theta)e^{+i\phi}|1,-1>\\
\nonumber
\end{equation}

\begin{equation}
|\chi_{0}>=-\frac{1}{\sqrt{2}}sin \theta e^{(-i \phi)} |1,+1> +
cos\theta |1,0>+\frac{1}{\sqrt{2}}sin\theta e^{+i\phi}|1,-1>\\
\end{equation}
These are of course automatically also eigenstate of $H$ but while,
$|\chi_{0}>$ is an eigenvector of $H$ belonging to the eigenvalue
$-(\frac{e^{2}q_{0}Q}{12})\frac{2}{3}$, the vectors $|\chi_{+1}>$
and $|\chi_{-1}>$ are both eigenvectors of $H$ belonging to the
eigenvalue $(\frac{e^{2}q_{0}Q}{12})\frac{1}{3}$, and hence we have
here a two fold degeneracy. While for the state $|\chi_{0}>$, the
Berry phase maybe obtained as earlier, since the instantaneous
eigenstates $|\chi_{+1}>$ and $|\chi_{-1}>$ are degenerate, it is
unclear whether true adiabatic change is possible for these two
states. However, since the change being made is through rotation
about the Z-axis, the states, which are eigenstates of $J_{z}$, are
the proper linear combination in the manifold of degenerate states,
for which the phase matrix is diagonalized, and consequently the
notion of Berry's phase remains meaningful, as discussed by
Wilczek and Zee, and Simon (ref. \cite{wilzchekpaper},\cite{simonpaper}) .\\
However, $J_{z}$ can be expressed as a linear combination of
$J_{x^{'}}$,$J_{y^{'}}$ and $J_{z^{'}}$ and hence of
$J_{+^{'}}$,$J_{-^{'}}$ and $J_{z^{'}}$. But since these operators
can only cause a change in the projection quantum number $\Delta
m^{'}=0,\pm 1$, it is only in the Kramer's degenerate subspace $\pm
\frac{1}{2}$ that the non-abelian nature of the underlying gauge
group manifests itself in a non-trivial manner. Indeed, in the
degenerate subspaces, $\pm m$ other than the case $m=\frac{1}{2}$,
the eigenstates of $J_{z^{'}}$ are automatically also the eigenstates of $J_{z}$.\\

The Berry phases for a spin $1$ particle in an electric field with a
rotating field gradient is given as follows-

\begin{tabular}{|c|c|c|}
\hline
\\
Eigenstates & Energy Eigenvalues & Eigenphase\\
\hline
\hline
\\
$|\chi_{+}>$ & $(\frac{e^{2}q_{0}Q}{12})\frac{1}{3}$ & $2\pi
cos\theta$\\
\hline
\\
$|\chi_{-}>$ & $(\frac{e^{2}q_{0}Q}{12})\frac{1}{3}$ & $-2\pi
cos\theta$\\
\hline
\\
$|\chi_{0}>$ & $-(\frac{e^{2}q_{0}Q}{12})\frac{2}{3}$ &
$\pi(1-cos\theta)$\\
\hline\\
\end{tabular}

\section{Conclusion}
In this exercise we have tried to investigate the effect of a
time dependent Hamiltonian for two particle systems,
concentrating on time dependent magnetic field effects on the Berry
phase of two coupled spin half particles and one spin one-half and spin
one particle coupled together. Such a study gave us an opportunity
to go back to the original work of Berry and discover that the
statement made by Berry regarding the phase of the wavefunction of a  spin one-half particle in a
rotating magnetic field has a general validity even for systems with
two particles. Hence we might be able to hypothesize that the
formula $2\pi m (1-cos\theta)$ will be correct for systems with
greater number of particles with different values of spin, where in
those cases, $m$ will refer to the value of $S_{z}$ component along
the direction of the magnetic field, $S_{z}$ being the $z$ component
of the total angular momentum of the particles. \\
In our quest to broaden the range of physical systems where the Berry phase may manifest itself,
we have considered electric field analogues of the above situation. Here, because of the Kramer's  degeneracy,
among states of angular momentum projection ($\pm m$), stemming from time-reversal invariance, Wilczek and Zee pointed out
the relevance of a non-abelian gauge field underlying these phases. This has led to an experimental verification in the case of the
levels of a nucleus of spin $\frac{3}{2}$  in the electric field gradients due to the crystalline field of a rotating solid sample (ref.\cite{tyckopaper}).
We discussed the generalized scenario starting from the simplest $j=1$ situation.\\
For both time varying magnetic and electric fields, we point out the
gauge variety of systems involving a vast range of timescales and
the use of a diversity of techniques and possibly several amusing
manifestations of the Berry phase posing challenges to experimental
technology. Lastly it is important to underline the fact that in the
kind of problems being brought forth in the point of view we
advocate, those involved in the very basis of quantum
mechanics,physics of atoms and molecules, solids, nuclei and
elementary particles can share a common platform of investigation.

\end{document}